\documentclass{optica-article}

\journal{opticajournal} 

\articletype{Research Article}

\usepackage{lineno}
\usepackage{physics}
\usepackage{cleveref}
\usepackage{hyperref}
\usepackage{wallpaper}
\usepackage{tabularx} 
\usepackage{hyperref}

\begin{document}

\title{Optical quantum teleportation with known amplitude distorting factors of teleported qubits}

\author{Mikhail S.Podoshvedov,\authormark{1,2} Sergey A. Podoshvedov,\authormark{1,2,*}}

\address{\authormark{1}Laboratory of quantum information processing and quantum computing, South Ural State University (SUSU), Lenin Av. 76, Chelyabinsk, Russia\\
\authormark{2}Laboratory of quantum engineering of light, South Ural State University (SUSU), Lenin Av. 76, Chelyabinsk, Russia}

\email{\authormark{*}sapodo68@gmail.com}

\begin{abstract*} 
We develop a quantum teleportation protocol of an unknown optical single rail qubit using a hybrid quantum channel composed of continuous variable (CV) states of certain parity. The quantum channel is characterized by two parameters: a squeezing parameter of single-mode squeezed vacuum (SMSV) state and the beam splitter (BS) parameter used to implement it. The CV part of the hybrid state belongs to Alice, while discrete variable (DV) half is controlled by Bob. The third parameter of the protocol is a parameter of the beam splitter, used to mix the CV components of the hybrid quantum state with unknown optical single-rail qubit. Even though the number of measurement results Alice sends may increase, Bob can obtain the original qubit half the time with an appropriate choice of parameter values. In almost half the remaining cases, Bob obtains the original qubit with distorted amplitudes, and both participants know the value of the distortion factors. This means that as the amount of classical information transmitted by Alice increases, they both gain greater access to partial information about the unitary transformations that the teleported qubits undergo, allowing Bob to continue using them or attempt to recover them to improve the protocol's efficiency. The proposed method is a generalization of quantum teleportation with a nonlocal photon used as a quantum channel and unknown single-rail optical qubit.     

\end{abstract*}

\section{Introduction}
Quantum teleportation is a method for transmitting an unknown quantum state of a particle to distant particle, without sending the original particle itself, using quantum entanglement and two bits of classical information \cite{1}. Initially, this quantum protocol was originally applied to two-level quantum systems and included two remote participants, Alice and Bob, who wished to transmit/receive the quantum information. Such a transfer requires the resource of quantum entanglement \cite{2,3}, which is initially distributed between the participants \cite{1}. Typically, Alice detects Bell states \cite{2} through a joint quantum measurement, and her state collapses as a result of the measurement, while Bob's qubit is simultaneously projected onto one of the states requiring some adjustment \cite{4}. Remote reconstruction of the output qubit requires additional classical information about what corrective action needs to be applied in accordance with the measurement result, and therefore it excludes superluminal communication. If Alice knows the parameters of her qubit, then quantum teleportation is reduced to remote state preparation \cite{5}.  
\par Besides its conceptual significance, the quantum teleportation protocol plays an important role in the development of practical applications of quantum information science \cite{6,7,8}. Among the technological solutions arising from the quantum teleportation protocol, it is worth mentioning quantum repeaters \cite{9}, measurement-based quantum computing \cite{10} and quantum gate teleportation \cite{11}. Currently, optical quantum teleportation is implemented in various ways including its realization with polarization photons \cite{12,13,14}, optical qubits in single rail \cite{15,16} and dual rail \cite{17,18} representations and in time-bin interpretation \cite{19}. It is interesting to observe how far an unknown qubit can be teleported \cite{20,21}. One of possible modifications of the quantum teleportation involves a quantum swapping protocol \cite{22,23}, in which the participants do not initially interact with each other, but obtain an entangled state through the measurement of Bell states by a third participant. Quantum teleportation is the basis for manipulating unitary transformations achieved by using special entangled state, performing a local measurement, and applying single-qubit operations \cite{7,11}. This approach underlies linear-optical quantum computing \cite{24,25}.
\par Quantum teleportation can be extended to quantum systems with dimensions greater than two, and in general we can talk about dividing the systems used into finite (discrete variable (DV) systems) and infinite known as continuous variable (CV) systems. In the standard representation, a two-mode squeezed vacuum state (TMSV) is used as the entanglement resource, and the Bell CV measurement based on homodyne detection is applied to the two modes. In the final step, Alice communicates her classical results to Bob, after which he can apply the corresponding displacement to his qubit \cite{26,27,28}. CV teleportation cannot achieve unit fidelity due to finite squeezing, and to improve the efficiency of the process, CV components can be combined with DV components to create a hybrid entanglement resource \cite{29,30,31,32,33,34,35}. CV components associated with photonic states can improve the output characteristics of the teleported qubit \cite{36,37,38,39}. 
\par In this paper, we develop a new protocol for quantum information teleportation of an unknown single rail qubit using a hybrid quantum channel. The parity entanglement state is half owned by Alice (CV part) and half by Bob (DV part). The quantum channel is described by two independent parameters. Mixing of the CV and DV states occurs on a beam splitter with arbitrary transmission and reflection. Therefore, the quantum teleportation of an unknown single-rail qubit can be controlled using three independent parameters, allowing the participants to control the precision of the output qubits at Bob's disposal. Quantum teleportation occurs after Alice measures the number of photons in her measurement modes and sends the measurement results to Bob to correct his output qubit. After selecting the appropriate values of the quantum channel parameters and BS parameter, Bob receives the original qubit half the time, and the qubit with the distorted amplitude the other half. Moreover, both Alice and Bob know the values of these amplitude distortion factors, allowing Bob to use them at his discretion or try to get rid of the amplitude distortion multiplier. In order for Alice to have full access to information about the results of her measurements, it is necessary to use efficient photon number resolving (PNR) detectors. Genuine photon number resolution can be achieved using a superconducting transition edge sensor (TES) maintained near the transition temperature \cite{40,41}. Overall, this approach generalizes the quantum teleportation of a single-rail qubit using a nonlocal photon and expands the capabilities of the technology. The approach used to implement quantum teleportation of an unknown single-rail qubit is as feasible as the method for creating the corresponding hybrid entangled state \cite{34}.

\section{Quantum teleportation with output original and amplitude distorted qubits}

\begin{figure}[htbp]
    \centering\includegraphics[width=0.9\textwidth]{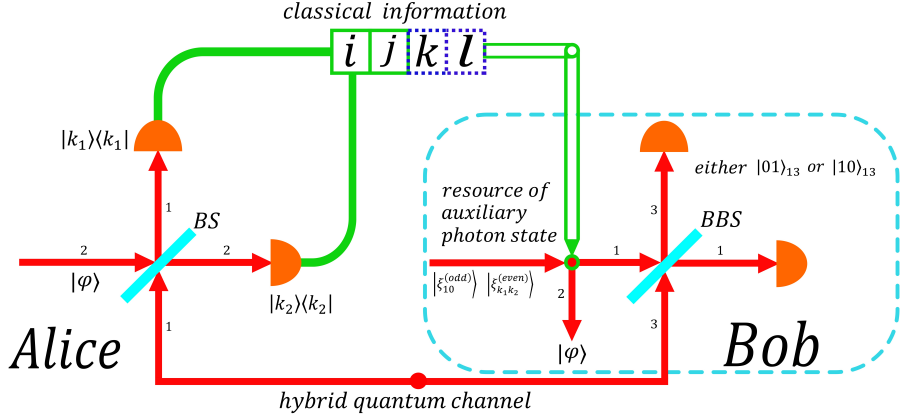}
    \caption{Schematic representation of the quantum teleportation protocol of photonic single-rail state \eqref{eq:2}  using optical hybrid quantum channel (1). The part of the quantum channel containing the CV states is mixed with the teleported qubit on balanced beam splitter, followed by detection of the measurement outcomes by two PNR detectors. Four measurement results 01,10,20,02 form the basis for Alice to compose two-bit messages (highlighted in green) for Bob. For certain values of the parameters of the quantum channel $S_{SMSV}\,\,(dB)$ and $B_0$, Bob receives the initial qubit with probability $P_{pt}=0.5$ provided that Alice measured 01 and 10. With the other measurement results from Alice, Bob receives a qubit with an amplitude distortion, i.e., the one on which the unitary transformation in equations (\ref{eq:10},\ref{eq:11}) has been performed. To help Bob determine which amplitude-distorted qubit he received, Alice should increase the number of bits kl transmitted (highlighted in dark blue). Given partial information about his qubit, Bob can attempt to reconstruct the original qubit using additional photon states (this part of the diagram is highlighted by the turquoise dotted line).}
    \label{Figure.1}
\end{figure}

Let us consider the following quantum hybrid channel in \hyperref[Figure.1]{Figure.~\ref*{Figure.1}} 

\begin{equation}\label{eq:1}
    \ket{\triangle_{1}}_{13} =
    \frac{1}{\sqrt{2}}\Bigl(
    \ket{\Psi_{1}^{(0)}}_1
    \ket{0}_3 +
    \ket{\Psi_{1}^{(1)}}_1
    \ket{1}_3\Bigl),
\end{equation}

\!\!\!\!\!\!connecting Alice (mode 1) and Bob (mode 3), who are located far apart. Here, the CV states with a certain parity $\ket{\Psi_{1}^{(0)}(y_0)}$ (odd CV state) and $\ket{\Psi_{1}^{(1)}(y_0,B_0)}$ (even CV state) are presented in the supplementary material. Their dependence on two initial parameters $y_0$ and $B_0$ is determined from the procedure of preparing the target quantum communication channel \cite{34}. 
The hybrid entangled state (\ref{eq:1}) can be realized by mixing the SMSV state with an auxiliary state $\ket{\varsigma}_{23} = \ket{00}_{23} + r\ket{11}_{23}$ at the beam splitter and then measuring the number of photons in the measurement mode of the system. The two-mode squeezed vacuum (TMSV) state $\ket{TMSV(s_2)} =
\sum_{n=0}^{\infty} \tanh^n s_2\ket{nn}/\cosh s_2$  \cite{42} with small squeezing amplitude $s_2$ can approximate the state $\ket{\varsigma}$. In the case of $s_2=0.1$, the amplitude of the correlated four-photon state $\ket{22}$ is about 100 times smaller than the amplitude of the state $\ket{11}$, which allows us to neglect higher-order correlated terms in TMSV state and deal with the superposition of vacuum and correlated single photon states \cite{42}. The CV states $\ket{\Psi_{1}^{(0)}}$  and $\ket{\Psi_{1}^{(1)}}$  follow directly from the initial SMSV state by subtracting one photon (subscript 1), provided that zero (superscript (0)) and one photon (superscript (1)), were initially added to the SMSV state. The analytical form of the CV states $\ket{\Psi_{1}^{(0)}}$  and $\ket{\Psi_{1}^{(1)}}$  is presented in the supplementary material.

Since the CV states follow directly from the SMSV state, their squeezing parameter $y_0$ is directly related to the squeezing parameter $y_{SMSV}=\tanh s /2$
with the squeezing amplitude $s>0$ of the original SMSV state through a decreasing factor $t_0^2$, that is,\,$y_0=y_{SMSV} t_0^2$, where the real parameter $t_0>0$ is the transmittance of BS through which the quantum channel in \eqref{eq:1} is realized \cite{34,35}. 
In the case of a highly transmissive beam splitter with $t_0\approx1$, the assumption $y\approx y_{SMSV}$ is valid, which means that the squeezing parameter does not change during propagation through the BS. In terms of the beam splitter parameter $B_0=r_0^2⁄t_0^2$, the reduced squeezing parameter can be rewritten as $y_0=y_{SMSV}t_0^2=y_{SMSV}⁄(1+B_0)$ since transmission $T_0=t_0^2$ and reflection $R_0=r_0^2$ coefficients ($r_0>0$) can be expressed in terms of $B_0$ as $T_0=1⁄(1+B_0)$ and $R_0=B_0⁄(1+B_0)$.
In addition, to characterize the initial SMSV state from which the quantum channel is created, one can use the squeezing $S_{SMSV}=-10 \log(\exp(-2s))$ in decibels (dB) and the average number of photons $\langle n_{SMSV} \rangle = \sinh^2 s$. As for the normalization factors of the CV states that form the hybrid entangled state \eqref{eq:1} , they are determined by the analytical function $Z(y_1)=1⁄\sqrt{1-4y_1^2 }$) and its derivatives. So, the normalization factor of the CV state $\ket{\Psi_{1}^{(0)}}$ is 
$G_1^{(0)}(y_0,B_0) = Z^{(1)}(y_0)/y_0,$ and the normalizations coefficient for the CV state 
$\ket{\Psi_{1}^{(1)}}$ becomes 
$G_1^{(1)}(y_0,B_0) = Z(y_0) - 2B_0(y_0Z^{(1)}(y_0)) + B_0^2(y_0\frac{d}{dy_0})(y_0Z^{(1)}(y_0)),$
where $Z^{(1)}(y_0)$ means the derivative of the function $Z(y_0)$, i.e. $Z^{(1)}(y_0) = dZ(y_0)/dy_0$. If we follow the behavior of the CV states in the case of $S_{SMSV}\rightarrow0$, then the CV state $\ket{\Psi_{1}^{(1)}}$ is transformed into a vacuum state, and the CV state $\ket{\Psi_{1}^{(0)}}$ is converted to a single photon, i.e., $\ket{\Psi_{1}^{(1)}} = \ket{0}$ and
$\ket{\Psi_{1}^{(0)}} = \ket{1}$, when $S_{SMSV} = 0$. Then, in the case the hybrid quantum state becomes a nonlocal photon $\ket{\triangle_{1}} =
    \Bigl(
    \ket{10} +
    \ket{01}\Bigl)/\sqrt{2}.$

\begin{figure}[htbp]
    \centering\includegraphics[width=0.9\textwidth]{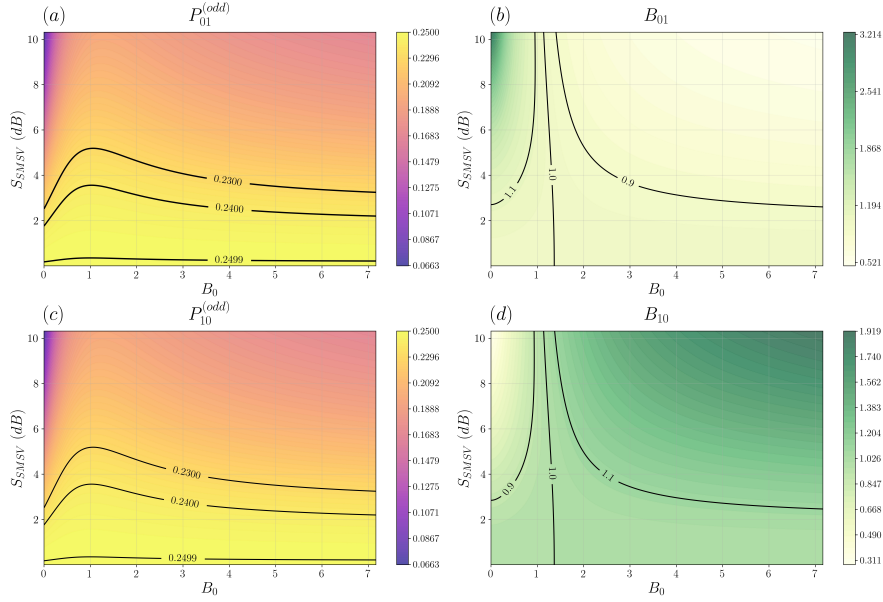}
    \caption{\!\!(a-d). (a) and (c) Contour plots of the probabilities $P_{01}^{(odd)}$ and $P_{10}^{(odd)}$ as functions of the quantum channel parameters $S_{SMSV}\,\,(dB)$ and $B_0$. In the selected range of $S_{SMSV}$ and $B_0$ variation, the probabilities can be monitored using a graduated color scale up to 0.25. The contour graph in (b) and (d) show dependences of the BS parameter $B_{01}$ and $B_{10}$, used in the quantum teleportation in \hyperref[Figure.1]{Fig.~\ref*{Figure.1}}, on the parameters $S_{SMSV}\,\,(dB)$ and $B_0$ in the same rectangle of their change as for (a) and (c). The straight lines $B=1$ on both graphs (b) and (d) are the same.}
    \label{Figure.2}
\end{figure}

The quantum channel in equation \eqref{eq:1}  is distributed between Alice and Bob as shown in \hyperref[Figure.2]{Figure.~\ref*{Figure.1}}. The CV state (mode 1) is directed toward Alice and the DV state (mode 3) is in Bob's hands. Alice also has an unknown single-rail qubit (mode 2)

\begin{equation}\label{eq:2}
    \ket{\varphi}_{2} =
    a_0
    \ket{0}_2 +
    a_1\ket{1}_2,
\end{equation}

\!\!\!\!\!\!which is a superposition of a vacuum and a single photon with corresponding unknown amplitudes $a_0$ and $a_1$ satisfying the normalization condition $\lvert{a_0}\rvert^2+\lvert{a_1}\rvert^2=1$. To teleport an unknown qubit, it is necessary to mix the CV part of the hybrid state with the unknown qubit on a beam splitter with real parameters $t>0$ and $r>0$, the matrix of which is presented in the supplementary material. It should be noted that the values of the parameters t and r can differ from $t_0$ and $r_0$ used to create the quantum channel. Thus, quantum teleportation of an unknown single-rail qubit is determined by three independent parameters, two of which $S_{SMSV}$ and $B_0$ are associated with quantum channel (1) and BS parameter $B=r^2⁄t^2$  already serves for the direct implementation of state mixing.
By changing these parameters, the maximum possible efficiency of this approach can be achieved. After CV-DV mixing on the BS in \hyperref[Figure.1]{Fig.~\ref*{Figure.1}}, whose parameters can be adjusted accordingly, modes 1 and 2 are measured using TES detectors \cite{40,41} capable of distinguishing the number of photons in each individual measurement channel. The interpretation of the quantum teleportation under consideration is reduced to the quantum teleportation of a single-rail photon by means of a nonlocal photon \cite{15,16} in the case of $S_{SMSV}=0$.
\par The conditional state that ends up in Bob's hands depends on the parity of the sum of the measurement outcomes in both measurement modes, i.e. $k_1+k_2$, where $k_1$ and $k_2$ are the number of measured photons in the first and second measurement modes. If this amount is an odd number, i.e., $k_1+k_2=2l+1$, then the state in Bob's hands becomes (removing the subscript 3 indicating the third mode in the initial quantum channel (1))

\begin{equation}\label{eq:3}
		|\varphi_{k_1 k_2}^{(\text{odd})}\rangle = \frac{a_0 |0\rangle + a_1b_{k_1 k_2}^{(\text{odd})} |1\rangle}{\sqrt{N_{k_1 k_2}^{(\text{odd})}}},
\end{equation}
which involves additional amplitude distortion multiplier  

\begin{equation}\label{eq:4}
		b_{k_1 k_2}^{(\text{odd})}(B) = \frac{1}{\sqrt{1+B}} \sqrt{\frac{Z^{(1)}(y_0)}{G_1^{(1)}(y_0,B_0)}} 
		\begin{cases}
			\frac{\sqrt{B(1+B)}}{y_0}\,\frac{(1-(k_1-1)B_0)}{2} , & \!\!\!\!\!\!\!\!\!\!\!\!\!\!\!\!\!\!\!\!\!
            \text{if } k_1 = 2l+1,\ l>0,\ k_2 = 0 \\[3ex]
			-\frac{k_2}{\sqrt{yB}}\frac{(1-(k_1+k_2-1)B_0)(1-\frac{k_1}{k_2}B)}{2(k_1+k_2)}, & \!\!\!\!\text{if } k_1+k_2 \text{ odd}
		\end{cases}
\end{equation}

\!\!\!\!\!\!which additionally also determines the normalization factor $N_{k_1 k_2}^{(odd)}=\lvert a_0 \rvert ^2+\lvert a_1 \rvert^2 b_{k_1 k_2}^{(odd)2}$ of the teleported qubit. The amplitude distortion factor is determined by two subscript $k_1$$k_2$ and superscript (odd). Using the mathematical conclusions that follow from this approach, one can derive the probability distribution of the measurement outcomes in the case of $k_1+k_2$ being odd

\begin{equation}\label{eq:5}
		P_{k_1 k_2}^{(\text{odd})}(y,B) = \frac{N_{k_1 k_2}^{(\text{odd})}}{2 Z^{(1)}(y_0)} \frac{y^{k_1} (yB)^{k_2}}{k_1! \, k_2!} \left( \frac{(k_1+k_2+1)!}{\left(\frac{k_1+k_2+1}{2}\right)!} \right)^2.
\end{equation}

\!\!\!\!\!\!Here, the parameter y is determined via the squeezing parameter $y_0$ of the quantum channel (1) by means of the reducing multiplier $t^2$, that is, $y=y_0t^2=y_0⁄(1+B)=y_{SMSV}⁄((1+B_0 )(1+B))$. As noted above, the BS parameter $B_0$ is used to create a quantum channel (1), and the BS parameter $B$ is used to implement the transmission of quantum information.
\par In the opposite case, when $k_1+k_2$ is an even number, the output state at Bob's disposal becomes as follows:

\begin{equation}\label{eq:6}
		|\varphi_{k_1 k_2}^{(\text{even})}\rangle = \frac{a_1 |0\rangle + a_0b_{k_1 k_2}^{(\text{even})} |1\rangle}{\sqrt{N_{k_1 k_2}^{(\text{even})}}},
\end{equation}

\!\!\!\!\!\!the amplitude distortion coefficient of which is determined by the formula

\begin{equation}\label{eq:7}
		b_{k_1 k_2}^{(\text{even})}(B) =  \sqrt{\frac{Z^{(1)}(y_0)}{G_1^{(1)}(y_0,B_0)}} 
		\begin{cases}
			\sqrt{\frac{y_0}{B}}\,\frac{1-k_1B_0}{k_1} , & \!\!\!\!\!\!\!\!\!\!\!\!\!\!\!\!\!\!\!\!\!
            \text{if } k_1 = 2l,\ l>0,\ k_2 = 0 \\[3ex]
			-\frac{\sqrt{y_0B}}{k_2}\frac{(1-(k_1+k_2)B_0)}{(1-\frac{k_1}{k_2}B)}, & \!\!\!\!\text{if } k_1+k_2 \text{ even}
		\end{cases}
\end{equation}

\!\!\!\!\!\!where the same notations are used as above, except that the superscript (odd) is replaced by (even) and the normalization factor is defined through this factor as $N_{k_1 k_2}^{(even)} =\rvert a_0 \lvert ^2+\rvert a_1 \lvert ^2 b_{k_1 k_2}^{(even)2}$. As follows from the supplementary material, the probability of the measurement outcomes $k_1 k_2$  is determined by the following distribution

\begin{equation}\label{eq:8}
		P_{k_1 k_2}^{(\text{even})}(y,B) =  \frac{N_{k_1k_2}^{(even)}}{2(1+B)Z^{(1)}(y_0)} 
		\begin{cases}
			B\frac{y^{k_1-1}}{k_1!}\Bigg(\frac{k_1!}{\Big(\frac{k_1!}{2}\Big)!}\Bigg)^2k_1^2, & \!\!\!\!\!\!\!\!\!\!\!\!\!\!\!\!\!\!\!\!\!\!\!\!\!\!\!\!\!\!\!\!\!\!\!\!\!\!\!\!\!\!\!\!\!\!\!\!\!\!\!\!\!\!\!\!\!\!\!\!\!\!\!\!\!\!\!\!\!\!\!\!\!\!\!\!\!\!\!\!\!\!\!\!\!\!\!\!\!\!\!
            \text{if } k_1 = 2l,\ l>0,\ k_2 = 0 \\[3ex]
			\frac{y^{k_1}(yB)^{k_2-1}}{k_1!k_2!}\Bigg(\frac{(k_1+k_2)!}{\Big(\frac{k_1+k_2}{2}\Big)!}\Bigg)^2\Big(1-\frac{k_1}{k_2}B\Big)^2k_2^2 , \, \text{if}\,\,k_1+k_2 \,\,even
		\end{cases}
\end{equation}

\par There remains one more measurement event $k_1=0$ and $k_2=0$, the registration of which does not allow Alice to transmit any superposition state. Instead of the DV superposition state, Bob will receive a single photon $\ket{1}$ with a probability $P_{00}^{(even)}(y,B)=\lvert a_0 \lvert^2⁄(2G_1^{(1)}(y_0,B_0))$ when $k_1=k_2=0$. Since the normalization factor $G_1^{(1)}(y_0,B_0)$ increases rapidly, this probability decreases quickly with increasing initial squeezing $S_{SMSV}$. We take the probability $P_{00}^{(even)}$ as the failure probability of the quantum teleportation protocol of an unknown optical DV qubit, i.e. $P_f=P_{00}^{(even)}$. 
\par We can now consider all possible measurement events to construct the output distribution of all possible measurement results. The overall distribution of measurement outcomes includes two partial distributions in equations (\ref{eq:5},\ref{eq:8}) and failure probability $P_f$

\begin{equation}\label{eq:9}
		P_{k_1 k_2}^{(\text{even})}(y,B) =   
		\begin{cases}
			\,\,\,\,\,\,\,\,\,\,P_f , & \!\!\!\!\!\!\!\!\!\!\!\!\!\!\!\!\!\!
            \text{if } k_1 = k_2 = 0 \\[3ex]
            P_{k_1k_2}^{(odd)}(y,B)
			, & \!\!\!\!\text{if } k_1+k_2\,\, odd \\[3ex]
            P_{k_1k_2}^{(even)}(y,B)
			, & \!\!\!\!\text{if } k_1+k_2 \,\,even
            
		\end{cases}
\end{equation}

By direct calculation one can verify that $P_{k_1 k_2} (y,B)$ is normalized that is, $\sum_{k_1=0}^{\infty}\sum_{k_2=0}^{\infty}P_{k_1k_2}(y,B) = P_f + \sum_{k_1=0}^{\infty}\sum_{k_2=0}^{\infty}\Big(P^{(odd)}_{k_1k_2}(y,B) + P^{(even)}_{k_1k_2}(y,B)\Big)=1$ for arbitrary values of y and B. 
Note that in some cases of the measurement outcomes, the amplitude distortion terms $b_{k_1 k_2}^{(odd)}(B)$ and $b_{k_1 k_2}^{(even)}(B)$ take on negative values. These negative values of the amplitude distortion factors follow from the expressions (\ref{eq:4},\ref{eq:7}). After receiving the corresponding measurement results, Bob will need to apply a phase shift operator to his qubit to get rid of the phase shift $\pi$.  
It follows from the above that the teleported state is determined by the parity of the sum of the measurement outcomes, which allows us to divide all possible measurement outcomes into two parts: those for which the sum $k_1+k_2$ is odd and those for which the sum $k_1+k_2$ is even. After receiving the measurement results ($k_1 k_2$ ), Alice knows exactly which state Bob has obtained, either in equation \eqref{eq:3}  or in \eqref{eq:6} . Moreover, knowing the values of the parameters $S_{SMSV}$, $B_0$ of the quantum channel (1) and BS parameter $B$, Alice also has complete information about the amplitude distortion term $b_{k_1 k_2}^{(odd/even)}$. If Alice knew the information about the initial amplitudes $a_0$ and $a_1$ of the qubit \eqref{eq:2} , she would know the unitary transformation of the input qubit to convert it to the output one

\begin{equation}\label{eq:10}
U^{odd}\begin{bmatrix}
        a_0 \\
        a_1 \\
      \end{bmatrix} = 
      \begin{bmatrix}
        U^{odd}_{11} & U^{odd}_{12}  \\
        U^{odd}_{21} & U^{odd}_{22}  \\
      \end{bmatrix}
      \begin{bmatrix}
        a_0 \\
        a_1 \\
      \end{bmatrix} = 
      \frac{1}{\sqrt{N^{(odd)}_{k_1k_2}}}
        \begin{bmatrix}
        a_0 \\
        b^{(odd)}_{k_1k_2}a_1 \\
      \end{bmatrix},      
\end{equation}

\begin{equation}\label{eq:11}
U^{even}\begin{bmatrix}
        a_0 \\
        a_1 \\
      \end{bmatrix} = 
      \begin{bmatrix}
        U^{even}_{11} & U^{even}_{12}  \\
        U^{even}_{21} & U^{even}_{22}  \\
      \end{bmatrix}
      \begin{bmatrix}
        a_0 \\
        a_1 \\
      \end{bmatrix} = 
      \frac{1}{\sqrt{N^{(even)}_{k_1k_2}}}
        \begin{bmatrix}
        b^{(even)}_{k_1k_2}a_1 \\
        a_0
      \end{bmatrix},      
\end{equation}

\!\!\!\!\!\!where matrix elements, for example, of $U^{(odd)}$ are given by

\begin{equation}
    U_{11}^{(odd)} = U_{22}^{(odd)} = \lvert a_0 \rvert^2 + \lvert a_1 \rvert^2 b^{(odd)}_{k_1k_2}, \nonumber     
\end{equation}

\begin{equation}\label{eq:12}
    U_{12}^{(odd)} = -U_{22}^{(odd)} = \lvert a_0 \rvert\lvert a_1 \rvert \exp(i\varphi) (1 - b^{(odd)}_{k_1k_2}),      
\end{equation}

\!\!\!\!\!\!where $\varphi$ is the azimuthal angle of the teleported qubit. Note that the transformation $U^{(even)}$ can be obtained as a product of the matrix $U^{(odd)}$ with the corresponding amplitude distortion factor  $b_{k_1 k_2}^{(even)}$ and a single qubit gate X, where X is the Pauli matrix, i.e. $U^{(even)}=U^{(odd)} \Big(b_{k_1 k_2}^{(even)}\Big)X$.
The inverse transformation of the output qubits (3,5) into the input qubit \eqref{eq:2}  can be implemented using unitary matrices $U^{(odd)-1}=U^{(odd)\dagger}$ and $U^{(even)-1}=U^{(even)\dagger}$ , where $U^\dagger$ is the operator Hermitian adjoint to the original U. In the quantum teleportation protocol, Alice does not know the exact values of $a_0$ and $a_1$, so she cannot calculate the matrix elements \eqref{eq:12}  and use the inverse transformation. Therefore, it can be said that the output state is subject to an unknown unitary transformation either \eqref{eq:10}  or \eqref{eq:11} , some information about which, related to the amplitude distortion coefficient, is known. 
\par In order to ensure that the output DV qubit is not subjected to the additional unitary transformations present in (\ref{eq:10},\ref{eq:11}), the protocol conditions must be chosen such that there is no amplitude distortion factor, i.e., that either $b_{k_1 k_2}^{(odd)}=1$ or $b_{k_1k_2}^{(even)}=1$, which leads to relation normalization factors $N_{k_1k_2}^{(odd)}=N_{k_1 k_2}^{(even)}=1$. 
To do this, Alice can vary the initial parameters $S_{SMSV}$,$B_0$ and $B$, which in some cases allows her to achieve the disappearance of amplitude distortion coefficients. Since the probabilities of the measurement outcomes with small values of $k_1$ and $k_2$ are higher than that with larger values of $k_1$ and $k_2$, it makes sense to consider in more detail special cases with $k_1+k_2=1$ and $k_1+k_2=2$. Therefore, we will limit ourselves here to two cases of odd sum $k_1+k_2=1$ and even sum $k_1+k_2=2$, which includes five combinations of the measurement outcomes, which we are going to denote as 01, 10, 02, 20 and 11. 
The amplitude distortion coefficients of the selected measurement results acquired by the output qubits follow from expressions (\ref{eq:4},\ref{eq:7})

\begin{equation}\label{eq:13}
		b_{01}^{(\text{odd})} = \frac{1}{2\sqrt{y_0 B}} \sqrt{\frac{Z^{(1)}(y_0)}{G_1^{(1)}(y_0,B_0)}},
\end{equation}

\begin{equation}\label{eq:14}
		b_{10}^{(\text{odd})} = \frac{1}{2}\sqrt{\frac{B}{y_0}} \sqrt{\frac{Z^{(1)}(y_0)}{G_1^{(1)}(y_0,B_0)}} = B b_{01}^{(\text{odd})},
\end{equation}

\!\!\!\!\!\!for odd 01 and 10 outcome and

\begin{equation}\label{eq:15}
		b_{20}^{(\text{even})} = \sqrt{\frac{y_0}{B}}\frac{(1-2B_0)}{2} \sqrt{\frac{Z^{(1)}(y_0)}{G_1^{(1)}(y_0,B_0)}} = B b_{02}^{(\text{even})},
\end{equation}

\begin{equation}\label{eq:16}
		b_{02}^{(\text{even})} = \frac{\sqrt{y_0B}(1-2B_0)}{2} \sqrt{\frac{Z^{(1)}(y_0)}{G_1^{(1)}(y_0,B_0)}} ,
\end{equation}

\begin{equation}\label{eq:17}
		b_{11}^{(\text{even})} = \frac{\sqrt{y_0B}(1-2B_0)}{1-B} \sqrt{\frac{Z^{(1)}(y_0)}{G_1^{(1)}(y_0,B_0)}} 
\end{equation}

\!\!\!\!\!\!for even 20, 02 and 11 outcomes, respectively, where we removed the minus sign for the corresponding coefficients. Numerical analysis shows that it is impossible to choose such values of the quantum channel $S_{SMSV}$,\,$B_0$ and BS parameter $B$ of the beam splitter used in quantum teleportation of an unknown qubit in \hyperref[figure.1]{figure.~\ref*{Figure.1}} that would ensure the simultaneous disappearance of all five amplitude distortion factors, i.e. the fulfillment of the condition $\lvert b_{01}^{(odd)} \rvert=b_{10}^{(odd)}=\lvert b_{02}^{(even)} \rvert=\lvert b_{11}^{(even)}\rvert=b_{20}^{(even)} =1$. Given the simple linear relationship between the amplitude distortion factors $b_{01}^{(odd)}$ and $b_{10}^{(odd)}$ as well as $b_{02}^{(even)}$ and $b_{20}^{(even)}$ through the BS parameter $B$, it can be assumed that for some values of $ S_{SMSV},\,B_0$ and $B$ some of the amplitude distortion factors can disappear, which allows for perfect teleportation. In the case the normalization factors $N_{k_1 k_2}^{(odd)}=N_{k_1 k_2}^{(even)}=1$ cease to depend on the unknown amplitudes $a_0$ and $a_1$.

We begin our consideration with the measurement outcome 01, the probability of which is given by

\begin{equation}\label{eq:18}
		P_{01}^{(\text{odd})} = \frac{B_{01}(1-4y_0^2)^{3/2}}{2(1+B_{01})},
\end{equation}

\!\!\!\!\!\!provided that $\lvert b_{01}^{(odd)} \rvert=1$, where the BS parameter $B_{01}$      

\begin{equation}\label{eq:19}
		B_{01} = \frac{Z^{(1)}(y_0)}{4y_0G_1^{(1)}(y_0,B)},
\end{equation}

\!\!\!\!\!\!follows from the condition. Output state \eqref{eq:3}  with amplitude distortion factor $b_{10}^{(odd)}=1$ is realized with probability

\begin{equation}\label{eq:20}
		P_{10}^{(\text{odd})} = \frac{(1-4y_0^2)^{3/2}}{2(1+B_{10})},
\end{equation}

\!\!\!\!\!\!provided that a beam splitter with the BS parameter
\begin{equation}\label{eq:21}
		B_{10} = \frac{4y_0G_1^{(1)}(y_0,B)}{Z^{(1)}(y_0)}
\end{equation}

\!\!\!\!\!\!is used. Note that the BS parameters $B_{01}$ and $B_{10}$ are interconnected as $B_{10}=1⁄B_{01}$  in the case of the absence of the corresponding amplitude distortion factors, which is a direct consequence of the linear dependence of the amplitude distorting multipliers \ref{eq:13}  and \ref{eq:14} . This means that if $B_{01}=1$, then $B_{10}=1$, and moreover $\lvert b_{01}^{(odd)} \rvert =b_{10}^{(odd)} $.

\par \hyperref[Figure.2]{Figure.~\ref*{Figure.2}(a)}(a) shows the dependence of the probability $P_{01}^{(odd)}$ in coordinates $S_{SMSV}$ and $B_0$, presented in the form of contour colors. As can be seen from this graph, the maximum probability, close to 0.25, is observed over a significant part of used rectangle $S_{SMSV}$, $B_0$, and the probability value in other places of the figure does not differ much from this maximum value. Therefore, it can be concluded that obtaining the output pure state \eqref{eq:3}  with  a unit coefficient of amplitude distortion is realized with probability $P_{01}^{(odd)}=0.25$ for an appropriate choice of $S_{SMSV}$ and $B_0$, provided that the measurement outcome 01 is registered by Alice. The corresponding contour plot for the BS parameter $B_{01}$ that should be used in the quantum teleportation protocol in case of the outcome 01 is shown in \hyperref[Figure.2]{Figure.~\ref*{Figure.2}(b)} In the rectangle of quantum channel parameters $S_{SMSV}$ and $B_0$ used for quantum teleportation, the beam splitter parameter $B_01$ can take values from 0.52 to 3.2. Similar contour dependencies for $P_{10}^{(odd)}$ and  $B_{10}$ for the measurement event 10 are shown in \hyperref[Figure.2]{Figure.~\ref*{Figure.2}(c,d)}. Although the $P_{10}^{(odd)}$ dependence in \hyperref[Figure.2]{Figure.~\ref*{Figure.2}(c)} may be similar to the $P_{01}^{(odd)}$ dependency in \hyperref[Figure.2]{figure.~\ref*{Figure.2}(a)}, the behavior of BS parameters $B_{10}$ in \hyperref[Figure.2]{figure.~\ref*{Figure.2}(d)} and $B_{01}$ in \hyperref[Figure.2]{figure.~\ref*{Figure.2}(b)} is already different. This difference is especially noticeable near the line $B_{01}=B_{10}=1$, where $B_{01}$ and $B_{10}$ behave differently depending on whether they are viewed from the right or left of the line. Note that the straight line $B_{01}=1$ in \hyperref[Figure.2]{figure.~\ref*{Figure.2}(b)} is completely identical to the line $B_{10}=1$ in \hyperref[Figure.2]{figure.~\ref*{Figure.2}(d)}. Thus, in the case of using the balanced beam splitter with $B=1$, the output state at Bob's location becomes identical to the original teleported state \eqref{eq:2} , and the events 01 and 10 occur with probability $P_{pt}=0.5$ (here the subscript pt means perfect teleportation, that is, teleportation of an unknown qubit without an amplitude distorting factor) for an appropriate choice of $S_{SMSV}$ and $B_0$ lying at the line $B=1$. 
\par Let us consider two even measurement outcomes 02 and 20 whose probabilities follow from the distribution \eqref{eq:8}

\begin{equation}\label{eq:22}
		P_{20}^{(\text{even})} = \frac{B_{20}(1-4y_0^2)^{3/2}}{(1+B_{20})^2},
\end{equation}

\begin{equation}\label{eq:23}
		P_{02}^{(\text{even})} = \frac{B_{02}(1-4y_0^2)^{3/2}}{(1+B_{02})^2},
\end{equation}

\!\!\!\!\!\!provided that the beam splitters with BS parameters  

\begin{equation}\label{eq:24}
		B_{01} = \frac{y_0(1-2B_0)^2}{4}\frac{Z^{(1)}(y_0)}{G_1^{(1)}(y_0,B_0))},
\end{equation}

\begin{equation}\label{eq:25}
		B_{01} = \frac{4}{y_0(1-2B_0)^2}\frac{G_1^{(1)}(y_0,B_0)}{Z^{(1)}(y_0)},
\end{equation}

\!\!\!\!\!\!are used. The use of beam splitters with BS parameters in Eqs. (\ref{eq:24},\ref{eq:25}) allows one to teleport an unknown qubit with either $b_{20}^{(even)}=1$ or $\lvert b_{02}^{(even)} \rvert=1$, respectively. In case of $B_{20}=B_{02}=B$ the probabilities $P_{20}^{(even)}$ and $P_{02}^{(even)}$ are equal to each other, i.e., $P_{20}^{(even)}=P_{02}^{(even)}$. Furthermore, the BS parameters are interrelated by the relation $B_{20}=1⁄B_{02}$  as in the case of single-photon measurement outcomes, which is a consequence of the linear relationship between the amplitude distortion coefficients. 
\par
\begin{figure}[htbp]
    \centering\includegraphics[width=0.9\textwidth]{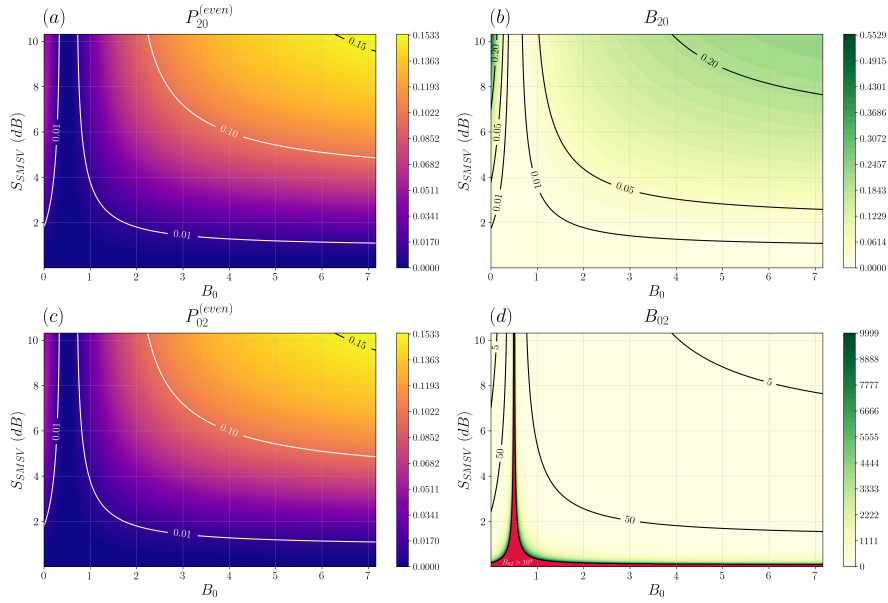}
    \caption{\!\!\!(a-d). (a) and (c) Contour plots of the probabilities $P_{20}^{(odd)}$ and $P_{02}^{(odd)}$ as functions of the parameters $S_{SMSV}\,\,(dB)$  and $B_0$ of the quantum channel. These dependencies are displayed in the same rectangle of $S_{SMSV}\,\,(dB)$ and $B_0$ as in the graphs in \hyperref[Figure.2]{figure.~\ref*{Figure.2}(b,d)}, the parameter $B_{20}$ takes values less than 1 ($B_{20}<1$), while $B_{02}$ becomes greater than one ($B_{02}>1$) to ensure elimination of the corresponding amplitude distorting factor. }
    \label{Figure.3}
\end{figure}

\hyperref[Figure.3]{Figures.~\ref*{Figure.3}(a,c)} show the corresponding contour dependencies of $P_{20}^{(even)}$ and $P_{02}^{(even)}$ on $S_{SMSV}$ and $B_0$. In contrast to the single photon dependencies, these probabilities take on smaller values and begin to increase to 0.25 only with the increase of the parameters of $S_{SMSV}$ and $B_0$. In general, the value of 0.25 is possible for $S>10\,\,dB$ and $B_0>7$. As for the beam splitter parameter $B$, for which an unity value of the amplitude-distortion factor is possible, the situation here differs from that which occurs when registering single photon. As shown in \hyperref[Figure.3]{Figure.~\ref*{Figure.3}(b)}, the measurement event 20 requires a beam splitter with $B_{20}<1$. On the contrary, to eliminate the amplitude distortion factor in the case of measurement even 02, a beam splitter with $B_{20}>1$ is required. Furthermore, it should be kept in mind that the output state becomes $\ket{\varphi_{20}^{(even)}}=\ket{\varphi_{02}^{(even)}}=a_1 \ket{0}+a_0\ket{1}$ even in the absence of the amplitude-distorting factor, which requires applying an additional bit-flip operation X to the qubit $X\ket{\varphi_{k_1k_2}^{(even)}} = \ket{\varphi}$, where X-gate is the Pauli matrix, to restore original state.
\par
\begin{figure}[htbp]
    \centering\includegraphics[width=0.7\textwidth]{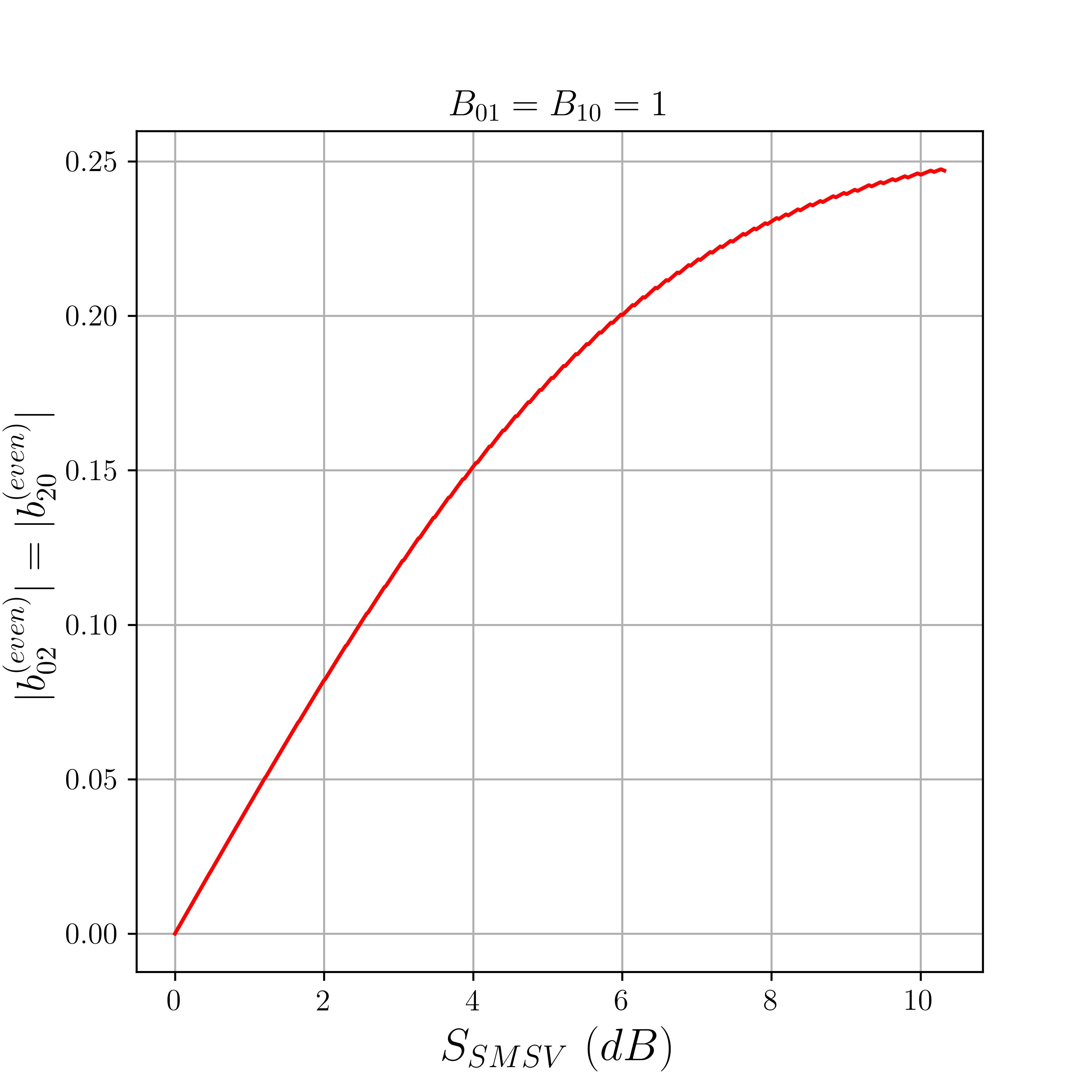}
    \caption{Dependence of amplitude distortion multipliers $b_{20}^{(even)}(B=1)=\lvert b_{02}^{(even)} (B=1)\rvert$  on the squeezing parameter $S_{SMSV}\,\,(dB)$ of the quantum channel (1). When constructing the curve, those values of $S_{SMSV}\,\,(dB)$ and $B_0$ are used that ensure a straight line $B=1$ in the \hyperref[Figure.2]{figures.~\ref*{Figure.2}(b,d)}. The amplitude distortion factors increase with increasing $S_{SMSV}\,\,(dB)$ but remain less than one.}
    \label{Figure.4}
\end{figure}

 As follows from the graphs in \hyperref[Figure.2]{Figures.~\ref*{Figure.2}} and \hyperref[Figure.3]{~\ref*{Figure.3}}, the conditions, under which the amplitude distortion multipliers for odd (single photon) and even (two photons) events are eliminated, differ for the same values of $S_{SMSV}$ and $B_0$ of the quantum channel (1), that is, they arise at different values of the BS parameter B. In case of a straight line $B=1$, the amplitude distortion factors $\lvert b_{02}^{(even)} \rvert $,\,$b_{20}^{(even)} $ are not equal to unity. This means that choosing $B=1$ allows teleporting an unknown qubit with probability 0.5, whereas measuring two-photon events (either 20 or 02 measurement outcomes) can lead to the output qubits already containing the amplitude distorting factors. However, the amplitude distortion coefficients are equal to each other $b_{20}^{(even)}=|b_{02}^{(even)}|\neq 1$ at $B=1$. \hyperref[Figure.4]{Figure.~\ref*{Figure.4}} shows the dependence of the amplitude-distorting multipliers on the squeezing $S_{SMSV}$ of the quantum channel  for those values of $B_0$ that provide the straight line $B=1$ in \hyperref[Figure.2]{Figure.~\ref*{Figure.2}(b,d)}. As can be seen from the figure, this dependence takes on values less than 1. The presence of the amplitude distorting factor does not allow us to estimate the probability of teleportation of the modified unknown qubit, the parameters of which $a_0$ and $a_1$ remain hidden from the participants.

\par Using the results of measurements (20) and (02)  as an example, it is worth noting the distinctive feature of the quantum teleportation of an unknown single-rail qubit using hybrid quantum channel (1). If quantum teleportation were performed using a quantum channel (1) with $S_{SMSV}=0$, then after receiving the measurement results (20) and (02), Bob would obtain a vacuum state, not a superposition state. Furthermore, knowing the value of the amplitude distortion factor, Bob can use such a qubit at his own discretion. This is also true for any arbitrary measurement outcome (ij), with the exception of result (00), the probability of which quickly tends to zero with increasing $S_{SMSV}$. Thus, we can conclude that this protocol is nearly deterministic, in which Bob obtains the initial state half the time, and in the other half of the cases, the qubit in Bob's hands obeys unitary information, so that some of this transformation information is known to both participants.
\par In addition to two-photon events 20 and 02, there is also an event of detecting two photons in two measuring channels, i.e. event 11, which can occur with a probability

\begin{equation}\label{eq:26}
		P_{11}^{(even)} = \frac{(1-B_{11})^2(1-4y_0^2)^{3/2}}{2(1+B_{11})^2}
	\end{equation}

\!\!\!\!\!\!provided that $\vert b_{11}^{(even)} \lvert=1$, a condition that enables to derive the corresponding BS parameter $B_{11}$. In the case of either $B_{11}\gg1$ or $B_{11}\ll1$ the probability can approach 0.5, i.e. $P_{11}^{(even)}\rightarrow 0.5$, if $y_0 \rightarrow 0$.  In the case of $B=1$, Bob receives a single photon, which is not in a superposition state with vacuum, with probability $P_{11}^{(even)}(B=1)=(\lvert a_0 \rvert^2 y_0^2 (1-2B_0 )^2 )⁄(2G_1^{(1)}(y_0,B_0))$ due to destructive photon interference. The probability $P_{11}^{(even)}(B=1)$ becomes zero in case of $S_{SMSV}=0$. The probability of the measurement outcome 11 for $B=1$ can be attributed to the probability of failure of the quantum teleportation protocol as well as the probability of event 00. For the values of $S_{SMSV}$ and $B_0$ used, the probability $P_{11}^{(even)}(B=1)$ also takes on small values. 

\par Thus, the quantum teleportation protocol of an unknown photonic single-rail qubit can be performed as follows:
\par 1. A hybrid entangled state \ref{eq:1}  with specific parameter values $S_{SMSV}$ and $B_0$ is prepared and distributed between Alice and Bob. Alice owns the CV fraction, and Bob has the DV part of the quantum channel at his disposal. Both participants know the values of the parameters $S_{SMSV}$ and $B_0$ in advance. Alice also decides to implement this protocol on a balanced beam splitter, which she informs Bob about in advance. This means that Alice chooses such values of $S_{SMSV}$ and $B_0$ so that they lie on the line $B=1$ in \hyperref[Figure.2]{figures.~\ref*{Figure.2}(b,d)}. This allows both participants to calculate the amplitude distortion coefficients of the teleported qubit before the protocol starts.
\par 2. Alice mixes the CV part of the hybrid entangled state \eqref{eq:1}  with an unknown single-rail qubit \eqref{eq:2}  on the balanced BS and measures the number of photons in the beam splitter's output modes. Having received the results of her measurements, Alice forwards this information to Bob via a classical communication channel. 
\par 3. If Bob receives 10 result , he does nothing. If he receives message 01, Bob applies Z-gate to his state to restore it. These two events occur with probability $P_{pt}=0.5$. If Alice sends Bob other measurement results, then, upon receiving messages 20 and 02, Bob knows that he has at his disposal the original qubit subjected to unitary transformations in equations (10,11), that is, a qubit that includes a pre-known amplitude distortion coefficient. Knowing the value of the amplitude distortion coefficient, Bob can continue to use the for its intended purpose. In the case of measurement result 00, the output qubit in Bob's hands is destroyed, and the probability of this outcome can be taken as the failure probability, which quickly tends to zero at the chosen value of $S_{SMSV}$. If Alice limits herself to four measurement results (01,10,20,02), then she sends Bob 2 bits of information. If Alice decides to send the measurements results of three photons (03,12,21,30), the classical information volume doubles from two to four bits. The measurement result 11 is discarded by Alice when she uses the beam splitter with $B=1$.    
\par 4. Overall, the proposed protocol of quantum teleportation of an unknown single-rail qubit is nearly deterministic, meaning it can be executed with probability close to 1, but not with unit fidelity. Half the time, Bob receives the original qubit, and the other half, the initial qubit, on which the unitary operation in Eqs. (\ref{eq:10},\ref{eq:11}) has been performed, and some information about this transformation is known to both participants in the protocol.

\section{Probabilistic reconstruction of photonic qubits with distorted amplitude}

Above, we considered the implementation of a quantum teleportation protocol for an unknown single-rail qubit from Alice to Bob via a quantum hybrid entangled state \eqref{eq:1} , where Bob does not attempt to correct his qubit, even knowing the amplitude distortion factor of his output qubit. In general, Bob can attempt to improve the efficiency of the quantum protocol by having partial knowledge of the output qubit, thereby eliminating the amplitude distortion factor and restoring the original qubit \eqref{eq:2} . Among the possible strategies for using unknown qubits with a known amplitude distortion coefficient, we will consider the most natural one. 
\par This approach involves mixing an amplitude-distorted qubit with a pre-prepared qubit in two modes on balanced beam splitter, followed by recording the measurement results of either 01 or 10. To implement the approach, Bob must first prepare auxiliary photon states

\begin{equation}\label{eq:27}
     \ket{\xi_{k_1k_2}^{(odd)}}_{13} = 
    \frac{1}{\sqrt{N_{k1k2}}}(\ket{01}_{13} + b^{(odd)}_{k_1k_2}\ket{10}_{13}),  
\end{equation}

\begin{equation}\label{eq:28}
     \ket{\xi_{k_1k_2}^{(even)}}_{13} = 
    \frac{1}{\sqrt{N_{k1k2}}}(b^{(even)}_{k_1k_2}\ket{00}_{13} + \ket{11}_{13}),  
\end{equation}

\!\!\!\!\!\!where $N_{k_1 k_2}=1+b_{k_1 k_2}^{\text{(odd,even)}2}$ are their normalization, factors and the qubits are represented in Bob’s local modes. After receiving a message from Alice, Bob can use one of the auxiliary states corresponding to the received message. So in case of receiving messages 20 and 02, Bob must use either $\ket{\xi_{20}^{(even)}}$ or $\ket{\xi_{02}^{(even)}}$ states, respectively.   
By mixing the auxiliary state with the teleported one, either $\ket{\varphi_{k_1k_2}^{(odd)}}$ or $\ket{\varphi_{k_1k_2}^{(even)}}$, located in the second mode at Bob's disposal, on the balanced beam splitter, Bob can restore the original state \eqref{eq:2}  provided that only one of the two detectors registers a single photon. The corresponding additional scheme at Bob's location, designed to eliminate the amplitude-distorting multipliers, is shown in \hyperref[Figure.1]{Figure.~\ref*{Figure.1}} (the dotted part of the figure), which Bob can use at his discretion. Probabilistic reconstruction allows us to estimate the overall probability of the quantum teleportation protocol with the recovery as 

\begin{equation}\label{eq:29}
     P_{pt} = 0.5 + P_{20}+ P_{02}+ P_{03}+ P_{12}+ P_{21}+ P_{30},  
\end{equation}

\!\!\!\!\!\!where partial even probabilities are given by

\begin{equation}\label{eq:30}
     P_{k_1k_2} = \frac{\lvert b_{k_1k_2}^{(even)}(B=1)\rvert ^2}{1 + \lvert b_{k_1k_2}^{(even)}(B=1)\rvert ^2}P_{k_1k_2}^{even}(B=1),  
\end{equation}

\!\!\!\!\!\!while odd partial probabilities are as follows

\begin{equation}\label{eq:31}
     P_{k_1k_2} = \frac{\lvert b_{k_1k_2}^{(odd)}(B=1)\rvert ^2}{1 + \lvert b_{k_1k_2}^{(odd)}(B=1)\rvert ^2}P_{k_1k_2}^{odd}(B=1),  
\end{equation}

\!\!\!\!\!\!Here we have taken into account odd probabilities with three photons from Eq. \eqref{eq:5}  and the corresponding photon distortion factors following from the expression \eqref{eq:4} . Since the quantum protocol is performed for the balanced BS, the value of the BS parameter takes on unity $B=1$ in the equations (\ref{eq:30},\ref{eq:31}). 
\par The probabilities in Eqs. (\ref{eq:30},\ref{eq:31}) no longer contain the normalization factor $N_{k_1 k_2}^{(odd,even)}$ with the unknown amplitudes of the teleported qubit. But the probabilities involve additional decreasing factor $\lvert b_{k_1k_2}^{(even,odd)}(B=1)\rvert^2⁄(1+\lvert b_{k_1k_2}^{(even,odd)}(B=1)\rvert ^2 )$, which can approach unity if $\lvert b_{k_1  k_2}^{(even,odd)}(B=1)\rvert \rightarrow 1$. This means that to increase the final probability, it is necessary for the amplitude distortion coefficients to take large values. Optimization calculations show that the teleportation probability can be increased to $P_{pt}=0,56$ provided that Bob applies the described recovery procedure, which includes three-photon measurement events. Note that if Bob receives other measurement outcomes 00,20,02 rather than 01 and 10, then the original qubit recovery protocol terminates, since Bob now loses control of his qubit. However, Bob can obtain amplitude distorted qubits when Alice measures four or more photons if Alice increases the number of transmitted bits.

\section{Conclusion}

Here we have proposed and explored a new protocol for quantum teleportation of an unknown optical single-rail qubit via a hybrid quantum channel \eqref{eq:1} . The quantum channel is formed from CV states of a certain parity, which are derivatives of the SMSV state, and photonic states, either vacuum or single photon. The CV components, as well as the DV states, are orthogonal to each other due to the difference in their parity. The quantum channel can be realized by mixing the SMSV state with the corresponding photonic state on a beam splitter and then measuring a certain number of photons in the measurement mode \cite{34}. The hybrid quantum channel is characterized by two parameters: the squeezing parameter $S_{SMSV}$ and the BS parameter $B_0$, which can be varied to control the output characteristics of the protocol.
The quantum teleportation protocol for an unknown photonic qubit is performed by mixing the CV part of the quantum channel with the unknown single-rail qubit on a balanced BS with $B=1$, and then measuring the number of photons in both modes at Alice's location. The obtained results are then forwarded to Bob via a classical communication channel to adjust the qubit at Bob's location accordingly. If Alice decides to limit herself to just four measurement results 01,10,20 and 02, she can do so by sending two bits of information. If Alice does not limit herself to just these measurement results, for example, if she decides to include three-photon measurement events, then she will need to formulate a message with more bits. Regardless of the number of measurement results sent by Alice (two or more), in half of the cases Bob will receive the original qubit after correction using Z-gate with an appropriate choice of quantum channel parameters. In the remaining cases, with the exception of measurement outcomes 00 and 11 with $B=1$, Bob receives the original qubit subject to amplitude distortion. However, given certain measurement outcomes, Alice and Bob know the value of the amplitude distortion factor, meaning they both have access to partial information about the unitary transformation the qubit may undergo during the teleportation. The probabilities of events 00 and 11 for $B=1$ are very small, allowing us to classify the proposed protocol for quantum teleportation of an unknown single-rail qubit as nearly deterministic, where Bob receives the original qubit half the time and a qubit with amplitude distortion, i.e., a qubit with non-unit fidelity, in the other half. From this point of view, the proposed version of quantum teleportation can resemble CV quantum teleportation with its non-perfect fidelity of the output state \cite{26}\cite{27}\cite{28}. Having partial information about the teleported qubit, Bob can use it for his own purposes, for example, for further teleportation of an already amplitude-distorted qubit. In particular, we considered the possibility of increasing the efficiency of the protocol in terms of probability by using additional auxiliary photon states created by Bob in advance.
\par In general, the proposed protocol is a generalization of quantum teleportation via the Bell state measurement. Indeed, the hybrid quantum channel \eqref{eq:1}  goes into a nonlocal photonic state  $\ket{\triangle_{1}} =\Bigl(\ket{10} + \ket{01}\Bigl)/\sqrt{2}$ in the case of $S_{SMSV}=0$. 
In the case, quantum teleportation of the unknown single-rail state \eqref{eq:2}  is accomplished using the nonlocal photon with probability 0.5 \cite{15,16}. In the remaining cases, Bob can only obtain either a vacuum state or a single photon, and further manipulations with output qubit, such as the recovery procedure, become impossible. The generalized version of quantum teleportation of the single-rail qubit with $S_{SMSV}\neq 0$ allows Bob to continue manipulating the output qubits, unlike its two-photon interpretation. The proposed protocol for quantum teleportation of an unknown single-rail qubit can be implemented using another hybrid quantum channel with a different ratio of the normalization coefficients, which is an important parameter of similar schemes. In the considered variant, the ratio $Z^{(1)}(y_0 )⁄G_1^{(1)}(y_0,B_0 )$ takes values less than 1 in the overwhelming majority of the rectangle $S_{SMSV}$ and $B_0$. This protocol can be scaled, which requires separate consideration. Both the implementation of quantum channel (1) and realization of the quantum teleportation protocol are feasible in practice.

\begin{backmatter}
    \bmsection{Acknowledgment}
     \!\!\!\!\!\!Supplementary material for the manuscript was prepared with the support of the Foundation for the Advancement of Theoretical Physics and Mathematics “BASIS” (Project № 24-1-1-87-1). 
    
    \bmsection{Disclosures}
    \!\!\!The authors declare no conflicts of interest.
    
\end{backmatter}

\renewcommand{\refname}{5.\,\,\,\,References}

\newpage

\title{Supplementary material for “Optical quantum teleportation with known amplitude distorting factors of teleported qubits”}

\author{Mikhail S.Podoshvedov,\authormark{1,2} Sergey A. Podoshvedov,\authormark{1,2,*}}

\address{\authormark{1}Laboratory of quantum information processing and quantum computing, South Ural State University (SUSU), Lenin Av. 76, Chelyabinsk, Russia\\
\authormark{2}Laboratory of quantum engineering of light, South Ural State University (SUSU), Lenin Av. 76, Chelyabinsk, Russia}

\email{\authormark{*}sapodo68@gmail.com}

\section*{1.\,CV states of definite parity used in quantum channel}

Here we have collected a collection of formulas related to the main text. The conclusions are based on a realistic theory of propagation of the SMSV state

\begin{equation}\label{eq:S1}
    \ket{SMSV(y)} =
    \frac{1}{\sqrt{\cosh{s}}}\sum_{n=0}^{\infty} \frac{y^{n}}{\sqrt{(2n)!}}\frac{(2n)!}{n!}\ket{2n},  
    \tag{S1}
\end{equation}

\!\!\!\!\!\!through a beam splitter (BS) with variable transmittance and reflection, which significantly increases its significance, since it allows obtaining accurate estimates of the statistical parameters for the BS with arbitrary parameters, i.e. 

\begin{equation}\label{eq:S2}
    BS_{ij} =
    \begin{bmatrix}
        t & -r  \\
        r & t  \\
      \end{bmatrix},
      \tag{S2}
\end{equation}

\!\!\!\!\!\!with arbitrary real transmittance $t>0$ and reflectance $r>0$  subject to the normalization condition $t^2+r^2=1$, where the subscripts $ij$ refer to the modes of the interacting light fields. 
\par As for the state of the SMSV state, the following designations are introduced here: squeezing parameter
$y = (\tanh s) /2$ is determined through 
the squeezing amplitude $s>0$, which imposes a limitation 
$0\leq y\leq0.5$ on it. In addition, the squeezing $S=-10\log(\exp(-2s)) \,\,dB$ in decibels and the average number of photons $\langle n_{SMSV} \rangle= \sinh^2 s$ can be used to characterize the original SMSV state.
\par The BS in equation \eqref{eq:S1}  mixes the modes 1 and 2 so that the creation operators $a_1^\dagger$ and $a_2^\dagger$ are transformed as $BS_{12}a_1^\dagger BS_{12}^\dagger=ta_1^\dagger-ra_2^\dagger$ and $BS_{12}a_2^\dagger BS_{12}^\dagger=ra_1^\dagger+ta_2^\dagger$, respectively. Applying the transformations to the initial SMSV state yields the output hybrid entangled state \cite{chap2:1,chap2:2}

\begin{equation}\label{eq:S3}
    BS_{12}\big(\ket{SMSV(y)}_1\ket{0}_2\big) = 
    \frac{1}{\sqrt{\cosh s}}\sum_{k=0}^{\infty} 
        \begin{array}{c}
            c_k^{(0)}(y_1,B)\sqrt{Z^{(k)}(y_1)}\ket{\Psi_k^{(0)}(y_1)}_1 \ket{k}_2
        \end{array} ,
    \tag{S3}
\end{equation}

\!\!\!\!\!\!which consists of continuous variable (CV) states of definite parity either even 

\begin{equation}\label{eq:S4}
    \ket{\Psi_{2m}^{(0)}(y_1)} =
    \frac{1}{\sqrt{Z^{(2m)}(y_1)}}\sum_{n=0}^{\infty} \frac{y_1^{n}}{\sqrt{(2n)!}}\frac{(2(n+m))!}{(n+m)!}\ket{2n}  
\tag{S4}
\end{equation}

\!\!\!\!\!\!for k=2m or odd

\begin{equation}\label{eq:S5}
    \ket{\Psi_{2m+1}^{(0)}(y_1)} =
    \sqrt{\frac{y_1}{{Z^{(2m+1)}}(y_1)}}\sum_{n=0}^{\infty} \frac{y_1^{n}}{\sqrt{(2n+1)!}}\frac{(2(n+m+1))!}{(n+m+1)!}\ket{2n+1}  
\tag{S5}
\end{equation}

\!\!\!\!\!\!for k=2m+1, while amplitudes of the output state \eqref{eq:S3}  are given by 

\begin{equation}\label{eq:S6}
    c_k^{(0)}(y_1,B) = (-1)^{k}\frac{(y_1B)^{\frac{k}{2}}}{\sqrt{k!}},
\tag{S6}
\end{equation}

\!\!\!\!\!\!which depend on the input squeezing parameter y decreased by $t^2$ times, that is, on $y_1=yt^2=y⁄(1+B)\leq y$, where BS parameter $B=(1-t^2)⁄t^2$  is introduced. Using the BS parameter $B$ one can determine its transmission $T=t^2=1⁄(1+B)$ and reflection coefficients $R=r^2=B⁄(1+B)$, respectively. From the definition it follows that the value $B=1$ corresponds to a balanced BS with T=R=0.5, while $B<1$ is responsible for transmitting beam splitter beam splitter with $T>R$ and the value $B>1$ describes a more reflective beam splitter with $R>T$. The normalization factors of the measurement induced CV states of definite parity (\ref{eq:S4},\ref{eq:S5}) are determined using derivatives of the analytical function $Z(y_1)=1⁄\sqrt{1-4y_1^2 }$ as $Z^{(2m)}(y_1)=dZ^{2m}⁄dy_1^{2m}$ and $Z^{(2m+1)}(y_1)=dZ^{(2m+1)}⁄dy_1^{(2m+1)}$, respectively. The subscript 2m,\,2m+1 is responsible for the number of subtracted photons, while the superscript takes into account the number of additional input photons, (0) corresponds to the input vacuum state.
\par The addition of input single photon modifies the basic model \eqref{eq:S3}  and leads to the generation of the following hybrid entangled state \cite{3}

\begin{equation}\label{eq:S7}
        \begin{array}{c}
         BS_{12}\big(\ket{SMSV(y)}_1\ket{1}_2\big) = (ra_1^{\dagger}+ta_2^{\dagger})BS_{12}\big(\ket{SMSV(y)}_1\ket{0}_2\big)= \\
            \frac{1}{\sqrt{\cosh s}}\sum_{k=0}^{\infty} c_k^{(1)}(y_1,B)
            \sqrt{G_k^{(1)}(y_1,B)}\ket{\Psi_k^{(1)}(y_1,B)}_1 \ket{k}_2,
        \end{array}
    \tag{S7}
\end{equation}

\!\!\!\!\!\!composed of new CV states of a certain parity

\begin{equation}\label{eq:S8}
    \ket{\Psi_{0}^{(1)}(y_1)} =
    \sqrt{\frac{y_1}{{G_{0}^{(1)}}(y_1)}}\sum_{n=0}^{\infty} \frac{y_1^{n}}{\sqrt{(2n+1)!}}\frac{(2n)!}{n!}(2n+1)\ket{2n+1},  
\tag{S8}
\end{equation}

\begin{equation}\label{eq:S9}
    \ket{\Psi_{2m}^{(1)}(y_1,B)} =
    \sqrt{\frac{y_1}{{G_{2m}^{(1)}}(y_1,B)}}\sum_{n=0}^{\infty} \frac{y_1^{n}}{\sqrt{(2n+1)!}}\frac{(2(n+m))!}{(n+m)!}\Bigg(1-\frac{2n+1}{2m}B \Bigg)\ket{2n+1},  
\tag{S9}
\end{equation}

\begin{equation}\label{eq:S10}
    \ket{\Psi_{2m+1}^{(1)}(y_1,B)} =
    \frac{1}{\sqrt{G_{2m+1}^{(1)}(y_1,B)}}\sum_{n=0}^{\infty} \frac{y_1^{n}}{\sqrt{(2n)!}}\frac{(2(n+m))!}{(n+m)!}\Bigg(1-\frac{2n}{2m+1}B \Bigg)\ket{2n},  
\tag{S10}
\end{equation}

\!\!\!\!\!\!with the following amplitudes \cite{3}

\begin{equation}\label{eq:S11}
    c_{k}^{(1)}(y_1,B) =\frac{1}{\sqrt{1 + B}}
    \begin{cases} \,\,\,\,\,\,\,\,\,\,\,\,\,\,\,\,\,\,\,\,\,\,\,\,\,\,\, \sqrt{B}, 
            \,\,\,\,\,\,\,\,\text{if}\,\,\, k = 0 \\[3ex]
			\,\,\,\,(-1)^{k+1}\frac{(y_1B)^{\frac{k-1}{2}}}{\sqrt{k!}}k , \,\,\text{if } k \neq 0  
	\end{cases}
\tag{S11}
\end{equation}

\par The new CV states \eqref{eq:S8}-\eqref{eq:S10}  have additional internal terms which complicate the normalization factors converting them to sum of the corresponding derivatives of the function $Z(y_1)$ as

\begin{equation}\label{eq:S12}
    G_{0}^{(1)}(y_1) = \frac{d}{dy_1}\Big(y_1Z(y_1)\Big) = Z^3(y_1) 
\tag{S12},
\end{equation}

\begin{equation}\label{eq:S13}
    G_{k}^{(1)}(y_1,B) = Z^{(k-1)}(y_1) + a_{k1}^{(1)}(B)(y_1Z^{(k)}(y_1)) + a_{k2}^{(1)}(B)y_1\frac{d}{dy_1}\Big(y_1Z^{(k)}(y_1)\Big), 
\tag{S13}
\end{equation}

\!\!\!\!\!\!with coefficients    

\begin{equation}\label{eq:S14}
   a_{k1}^{(1)}(B) = \frac{-2B}{k}, 
\tag{S14}
\end{equation}

\begin{equation}\label{eq:S15}
   a_{k2}^{(1)}(B) = \Big(\frac{B}{k}\Big)^2. 
\tag{S15}
\end{equation}

\!\!\!\!\!\!Here, for the coefficients of the corresponding derivatives of the function $Z(y_1)$ in equations (\ref{eq:S14},\,\ref{eq:S15}), the subscript k denotes the number of subtracted photons and the second number after k is the serial number of the coefficient, and the superscript (1), as above, is responsible for the number of added photons.

\section*{2.\,CV states used in quantum teleportation protocol  }

The hybrid entanglement (formula \eqref{eq:1}  in the main text) is constructed on the basis of the CV states of different parity for both CV states and photonic states as is described in \cite{2}. Here we only present the analytical form of the CV states that form the quantum channel in equation \eqref{eq:1}  of the main text. They are odd CV state

\begin{equation}\label{eq:S16}
    \ket{\Psi_{1}^{(0)}(y_1)} =
    \sqrt{\frac{y_1}{Z^{(1)}(y_1)}}\sum_{n=0}^{\infty} \frac{y_1^{n}}{\sqrt{(2n+1)!}}\frac{(2(n+1))!}{(n+1)!}\ket{2n+1},  
\tag{S16}
\end{equation}

\!\!\!\!\!\!and even CV state

\begin{equation}\label{eq:S17}
    \ket{\Psi_{1}^{(1)}(y_1)} =
    \frac{1}{\sqrt{G_1^{(1)}(y_1,B)}}\sum_{n=0}^{\infty} \frac{y_1^{n}}{\sqrt{(2n)!}}\frac{(2n)!}{n!}\Big(1-2nB\Big)\ket{2n},  
\tag{S17}
\end{equation}

\!\!\!\!\!\!whose the normalization factor becomes 

\begin{equation}\label{eq:S18}
    G_{k}^{(1)}(y_1,B) = Z^{(k-1)}(y_1) - 2B(y_1Z^{(k)}(y_1)) + B^2y_1\frac{d}{dy_1}\Big(y_1Z^{(k)}(y_1)\Big). 
\tag{S18}
\end{equation}

\par To implement the quantum protocol using a quantum optical channel in equation \eqref{eq:1}  of the main text, we need to additionally consider all possible CV states of a certain parity that are formed after the interaction of the CV states (\ref{eq:S16},\,\ref{eq:S17}) on BS with the vacuum and a single photon. In general, the system under consideration can be represented as a system of two beam splitters located one behind the other with BS parameters $B_0$ and $B$, respectively, through which the SMSV state passes. Additional single photons can be fed into the auxiliary modes of the system, or the inputs can be empty (vacuum state).
\par So, we have the following CV states of definite parity

\begin{equation}\label{eq:S19}
    \ket{\Psi_{1,2m_2}^{(00)}(y_2)} =
    \sqrt{\frac{y_2}{{Z^{(2m_2+1)}}(y_2)}}\sum_{n=0}^{\infty} \frac{y_2^{n}}{\sqrt{(2n+1)!}}\frac{(2(n+m_2+1))!}{(n+m_2+1)!}\ket{2n+1},  
\tag{S19}
\end{equation}

\!\!\!\!\!\!for $k_2=2m_2$ and

\begin{equation}\label{eq:20}
    \ket{\Psi_{1,2m_2+1}^{(00)}(y_2)} =
    \frac{1}{\sqrt{Z^{(2m_2+2)}(y_2)}}\sum_{n=0}^{\infty} \frac{y_2^{n}}{\sqrt{(2n)!}}\frac{(2(n+m_2+1))!}{(n+m_2+1)!}\ket{2n},  
\tag{S20}
\end{equation}

\!\!\!\!\!\!for $k_2=2m_2$. From here on, two superscripts are used (in the case (00)), which are responsible for the input photon states, and two subscripts (1$k_2$ ), which indicate the number of subtracted photons after the passing of the SMSV state through two BSs. Here the input squeezing parameter $y_1$ decreases by $t^2$ becoming $y_2=y_1⁄(1+B)=y⁄((1+B_0 )(1+B))$.  
\par Let's assume that the SMSV state interacts with the vacuum at the first beam splitter with BS parameter $B_0$ and with a single photon at the second beam splitter with BS parameter $B$, then we have the following set of new measurement induced CV states of definite parity

\begin{equation}\label{eq:S21}
    \ket{\Psi_{10}^{(01)}(y_2)} =
    \frac{1}{\sqrt{G_{10}^{(01)}(y_2)}}\sum_{n=0}^{\infty} \frac{y_2^{n}}{\sqrt{(2n+1)!}}\frac{(2n)!}{n!}2n\ket{2n},  
\tag{S21}
\end{equation}

\begin{equation}\label{eq:S22}
    \ket{\Psi_{1\,2m_2}^{(01)}(y_2,B)} =
    \frac{1}{\sqrt{G_{1\,2m_2}^{(01)}(y_2,B)}}\sum_{n=0}^{\infty} \frac{y_2^{n}}{\sqrt{(2n)!}}\frac{(2(n+m_2))!}{(n+m_2)!}\Big(1-\frac{n}{m_2}B\Big)\ket{2n},  
\tag{S22}
\end{equation}

\begin{equation}\label{eq:S23}
    \ket{\Psi_{1\,2m_2+1}^{(01)}(y_2,B)} =
    \sqrt{\frac{y_2}{G_{1\,2m_2+1}^{(01)}(y_2,B)}}\sum_{n=0}^{\infty} \frac{y_2^{n}}{\sqrt{(2n+1)!}}\frac{(2(n+m_2+1))!}{(n+m_2+1)!}\Big(1-\frac{2n+1}{2m_2+1}B\Big)\ket{2n+1},  
\tag{S23}
\end{equation}

\!\!\!\!\!\!where their normalizations factors are given by

\begin{equation}\label{eq:S24}
    G_{k_10}^{(01)}(y_2) = \Big(y_2\frac{d}{dy_2}\Big)\Big(y_2Z^{(k_1)}(y_2)\Big) 
\tag{S24},
\end{equation}

\begin{equation}\label{eq:S25}
    G_{k_1k_2}^{(01)}(y_2,B) = Z^{(k_1+k_2-1)}(y_2) + -\frac{2B}{k_2}y_2Z^{(k_1+k_2)}(y_2) + \frac{B^2}{k_2^2}\Big(y_2\frac{d}{dy_2}\Big)\Big(y_2Z^{(k_1+k_2)}(y_2)\Big). 
\tag{S25}
\end{equation}

\par Let's consider the case where a single photon is used in auxiliary mode of the first beam splitter with BS parameter $B_0$, but auxiliary mode of the second beam splitter with BS parameter $B$ is empty. Then, after simultaneous measurement of the number of photons ($1\,k_2$ ) in the output modes of both beam splitters, the following measurement induced CV states of a certain parity

\begin{equation}\label{eq:S26}
    \begin{array}{c}
         \ket{\Psi_{1\,2m_2}^{(10)}(y_2,B_0)} =
         \frac{1}{\sqrt{G_{1\,2m_2}^{(10)}(y_2,B_0)}}\sum_{n=0}^{\infty} \Bigg (\frac{y_2^{n}}{\sqrt{(2n)!}}
         \frac{(2(n+m_2))!}{(n+m_2)!}\\
         \Big(1-2m_2B_0-2nB_0\Big)\ket{2n}\Bigg),
    \end{array}
    \tag{S26}
\end{equation}

\begin{equation}\label{eq:S27}
    \begin{array}{c}
         \ket{\Psi_{1\,2m_2+1}^{(10)}(y_2,B_0)} =
         \sqrt{\frac{y_2}{G_{1\,2m_2+1}^{(10)}(y_2,B_0)}}\sum_{n=0}^{\infty} \Bigg (\frac{y_2^{n}}{\sqrt{(2n+1)!}}
         \frac{(2(n+m_2+1))!}{(n+m_2+1)!}\\
         \Big(1-(2m_2+2n+2)B_0\Big)\ket{2n+1}\Bigg),
    \end{array}
    \tag{S27}
\end{equation}

\!\!\!\!\!\!are realized, where their normalization factors are given by  

\begin{equation}\label{eq:S28}
    G_{1\,k_2}^{(10)}(y_2,B_0) = A^{(10)}_{1\,k_2\,0}Z^{(k_2)}(y_2) + 
    A^{(10)}_{1\,k_2\,1}y_2Z^{(k_2+1)}(y_2) + 
    A^{(10)}_{1\,k_2\,2}\Big(y_2\frac{d}{dy_2}\Big)\Big(y_2Z^{(k_2+1)}(y_2)\Big), 
\tag{S28}
\end{equation}

\!\!\!\!\!\!with amplitudes

\begin{equation}\label{eq:S29}
    A_{1\,k_2\,l}^{(10)}(y_1,B_0) =
    \begin{cases}
            \,\,\,\,\,\,\Big(1-k_2B_0\Big)^2,
            \,\,\text{if}\,\, l = 0 \\[3ex]
			\,\,\,-2\Big(1-k_2B_0\Big)B_0,\,\text{if} \,\,\,l = 1 .\\[3ex]
			\,\,\,\,\,\,B_0^2,\,
            \,\text{if} \,\,\,l = 2
	\end{cases}
\tag{S29}
\end{equation}

\par The fourth case involves the interaction of the SMSV state with two single photons on two beam splitters with parameters $B_0$ and $B$, respectively. After interaction, the measurement modes of the given system are simultaneously measured by two PNR detectors, resulting in new measurement induced CV states of a certain parity

\begin{equation}\label{eq:S30}
    \begin{array}{c}
         \ket{\Psi_{1\,0}^{(11)}(y_2,B_0)} =
         \sqrt{\frac{y_2}{G_{1\,0}^{(11)}(y_2,B_0)}}\sum_{n=0}^{\infty} \Bigg (\frac{y_2^{n}}{\sqrt{(2n+1)!}}
         \frac{(2n)!}{n!}(2n+1)\\
         \Big(1+B_0-B_0(2n+1)\Big)\Bigg)\ket{2n+1},
    \end{array}
    \tag{S30}
\end{equation}

\begin{equation}\label{eq:S31}
    \begin{array}{c}
         \ket{\Psi_{1\,2m_2}^{(11)}(y_2,B_0,B)} =
         \sqrt{\frac{y_2}{G_{1\,2m_2}^{(11)}(y_2,B_0,B)}}\sum_{n=0}^{\infty} \Bigg (\frac{y_2^{n}}{\sqrt{(2n+1)!}}
         \frac{(2(n+m_2))!}{(n+m_2)!}\\
         \Big(1-(2m_2-1)B_0-(2n+1)B_0\Big)\Big(1-\frac{B}{2m_2}(2n+1)\Big)\Bigg)\ket{2n+1},
    \end{array}
    \tag{S31}
\end{equation}

\begin{equation}\label{eq:S32}
    \begin{array}{c}
         \ket{\Psi_{1\,2m_2+1}^{(11)}(y_2,B_0,B)} =
         \frac{1}{\sqrt{G_{1\,2m_2+1}^{(11)}(y_2,B_0,B)}}\sum_{n=0}^{\infty} \Bigg (\frac{y_2^{n}}{\sqrt{(2n)!}}
         \frac{(2(n+m_2))!}{(n+m_2)!}\\
         \Big(1-2m_2B_0-B_0(2n)\Big)\Big(1-\frac{B}{2m_2+1}(2n)\Big)\Bigg)\ket{2n},
    \end{array}
    \tag{S32}
\end{equation}

\!\!\!\!\!\!where their normalization factors are given by

\begin{equation}\label{eq:S33}
\begin{array}{c}
         G_{k_1\,0}^{(11)}(y_2,B_0) = \Big(1+\frac{B_0}{k_1}\Big)^2\Big(y_2\frac{d}{dy_2}\Big)\Big(y_2Z^{(k_2-1)}(y_2)\Big)- \\
         2\Big(1+\frac{B_0}{k_1}\Big)\frac{B_0}{k_1}\Big(y_2\frac{d}{dy_2}\Big)^2\Big(y_2Z^{(k_2-1)}(y_2)\Big)+
         \Big(\frac{B_0}{k_1}\Big)^2\Big(y_2\frac{d}{dy_2}\Big)^3\Big(y_2Z^{(k_2-1)}(y_2)\Big),
    \end{array}
\tag{S33}
\end{equation}

\begin{equation}\label{eq:S34}
\begin{array}{c}
         G_{1\,k_2}^{(11)}(y_2) = A^{(11)}_{1\,k_2\,0}Z^{(k_2-1)}(y_2) + \sum_{l=1}^{4}A^{(11)}_{1\,k_2\,l}
        \Big(y_2\frac{d}{dy_2}\Big)^{l-1}\Big(y_2Z^{(k_2)}(y_2)\Big) , 
    \end{array}
\tag{S34}
\end{equation}

where the coefficients $A_{1\,k_2\,l}^{(11)}$ are formed from the internal amplitudes of the measurement induced CV states of a certain parity 

\begin{equation}\label{eq:S35}
a_{1\,k_2\,0}^{(11)} = 1 - (k_2-1)B_0,\,\,\,a_{1\,k_2\,1}^{(11)} = \Big(\frac{(k_2-1)B-1-k_2}{k_2}\Big)B_0,\,\,\,
a_{1\,k_2\,2}^{(11)} = \frac{B_0B}{k_2}
\tag{S35}
\end{equation}

\begin{equation}\label{eq:S36}
    \begin{array}{c}
        A^{(11)}_{1\,k_2\,0} = a^{(11)2}_{1\,k_2\,0},\,\,A^{(11)}_{1\,k_2\,1} = 2a^{(11)}_{1\,k_2\,0}a^{(11)}_{1\,k_2\,1},\,\,
        A^{(11)}_{1\,k_2\,2} = a^{(11)2}_{1\,k_2\,1}+2a^{(11)}_{1\,k_2\,0}a^{(11)}_{1\,k_2\,2}\\
        A^{(11)}_{1\,k_2\,3} = 2a^{(11)}_{1\,k_2\,1}a^{(11)}_{1\,k_2\,2},\,\,
        A^{(11)}_{1\,k_2\,4} = a^{(11)2}_{1\,k_2\,2}.
    \end{array}
\tag{S36}
\end{equation}

The given CV states form the basis for the development of a mathematical apparatus for quantum teleportation of an unknown single-rail qubit using a hybrid quantum channel (formula \eqref{eq:1}  in main text). Using this set of the CV states of a certain parity, we can find the output hybrid state in the case of mixing CV part of the quantum channel with single-rail qubit (formula \eqref{eq:2}  in the main text) $BS_{12}(\ket{\triangle_1}_{13}\ket{\varphi}_{2})$. The number of photons is measured in two modes of the hybrid entangled state (modes 1 and 2) and the simultaneous measurement of number of photons generates one of the two states: either $\ket{\varphi_{k_1k_2}^{(odd)}}$ (formula \eqref{eq:3}  in main text) or $\ket{\varphi_{k_1k_2}^{(even)}}$ (formula \eqref{eq:6}  in main text), depending on parity of the sum of the measured photon numbers. The amplitude distorting coefficients arise from the mismatch of the amplitudes of the hybrid entangled state, i.e., from the fact that, in general, the amplitudes $c_k^{(0)}(y_1,B)$ in Eq. \eqref{eq:S6}  are not equal to $c_k^{(1)}(y_1,B)$ in Eq. \eqref{eq:S11} . The amplitude distorting multipliers are proportional to the ratio $c_k^{(1)}(y_1,B)⁄c_k^{(0)}(y_1,B) )$ in the case of an odd sum $k_1+k_2$ of the measurement outcomes and $c_k^{(0)}(y_1,B)⁄c_k^{(1)}(y_1,B)$ in the case of an even sum $k_1+k_2$. The probabilities of measurement outcomes with measurement numbers $k_1$ and $k_2$ also follow from the amplitudes $c_k^{(0)}(y_1,B)$ and $c_k^{(1)}(y_1,B)$.

\section*{Acknowledgments}
The work of MSP and SAP was supported by the Foundation for the Advancement of Theoretical Physics and Mathematics “BASIS” (Project № 24-1-1-87-1).

\renewcommand{\refname}{References}

\end{document}